\documentclass[%
 reprint,
 amsmath,amssymb,
 aps,
]{revtex4-1}

\usepackage{graphicx}
\usepackage{dcolumn}
\usepackage{bm}
\usepackage{xcolor}
\usepackage{soul} 

\begin{document}


\title{Heavy quark mass effects in parton-to-kaon hadronization probabilities.}
\author{Manuel Epele}\email{manuepele@fisica.unlp.edu.ar}
\affiliation{ Instituto de F\'{\i}sica La Plata, UNLP, CONICET 
Departamento de F\'{\i}sica,  Facultad de Ciencias Exactas, Universidad de
La Plata, C.C. 69, La Plata, Argentina}
\author{Carlos Garc\'{\i}a Canal}\email{garcia@fisica.unlp.edu.ar }
\affiliation{ Instituto de F\'{\i}sica La Plata, UNLP, CONICET 
Departamento de F\'{\i}sica,  Facultad de Ciencias Exactas, Universidad de
La Plata, C.C. 69, La Plata, Argentina}
\author{R. Sassot}\email{sassot@df.uba.ar} 
\affiliation{Departamento de F\'{\i}sica and IFIBA,  Facultad de Ciencias Exactas y
 Naturales, Universidad de Buenos Aires, Ciudad Universitaria, Pabell\'on\ 1 (1428) 
Buenos Aires,  Argentina}
\begin{abstract}
We examine the relevance of the heavy quarks masses in the perturbative QCD description of hard interactions where charged kaons are produced in the final state. We extract a set of parton-to-kaon hadronization probabilities from a next to leading order QCD global analysis where a general mass variable flavor number scheme accounting for mass effects is implemented. We compare the results with those obtained in the massless approximation and also with those found in the case of final state pions. At variance with the very significant improvement found for the much more precise pion fragmentation phenomenology, the heavy quark mass dependent scheme improves mildly the overall description of current kaon production data. Nevertheless, the changes in the charm hadronization probability are noticeable.
\end{abstract}

\pacs{Valid PACS appear here}
\maketitle


\section{\label{sec:level1} Introduction}
Ultra-relativistic collisions produce very large numbers of particles with transverse momentum of several GeV. The most sought signals, those whose behaviour deviates from our present paradigm and that could indicate novel physical phenomena, are expected to be hidden beneath ordinary events which constitute an overwhelming background. The largest fraction of this background in experiments such as those performed at the Large Hadron Collider \cite{Abelev:2013ala} and at the Relativistic Heavy Ion Collider \cite{Agakishiev:2011dc} are light hadrons, such as pions and kaons. These are produced in the final state through the hadronization or fragmentation mechanism by which hard interacting partons evolve into a physical and intrinsically non-perturbative colorless hadronic state. In the context of perturbative Quantum Chromodynamics (QCD) \cite{Collins:1989gx}, hard hadronic collisions with identified final state hadrons are described perturbatively in terms of effective hard scattering cross sections and two sets of universal non-perturbative functions: parton distributions (PDF), which describe the internal structure of the hadrons just before the interaction process, and fragmentation functions (FF) that encode the information about the hadronization processes \cite{Feynman:1973xc}. 

Because of their non perturbative nature, PDFs and FFs need to be extracted through QCD global analyses of experimental data, where the hard scattering cross sections are approximated with increasing precision \cite{Butterworth:2015oua,Metz:2016swz}. The first generations of these global analyses relied on the massless quark approximations of QCD but progressively, a growing interest has been focussed in how the non-perturbative distributions are affected by considering quarks as massive particles. Of course, the relevance of dynamical effects associated to the quarks masses in a hard interaction depends crucially on both the masses and the energy scale that characterizes the process. For {\it up}, {\it down} and {\it strange} quarks, the same restriction that allows a perturbative treatment, i.e. energy scales much larger than $\Lambda_\text{QCD}$, guarantees the smallness of potential dynamical effects arising from their masses, making natural to treat them as massless, with the advantage of the all-order resummations implicit in massless parton approaches. However, this is not the case for {\it charm} and {\it bottom} quarks, whose mass thresholds fall inside the perturbation domain and produce the corresponding dynamical signatures and consequently need all-order resummations at very high energy scales. The so called general mass factorization schemes with a variable number of flavors (GMVFN) reproduce accurately both the massive and the massless regimes, smoothly interpolating between them \cite{Aivazis:1993pi}.

In the case of PDFs, the implementation of different variants of a GMVFN factorization scheme has become the standard practice to include heavy quark mass effects, keeping the consistency to the high energy limit \cite{Butterworth:2015oua}. The strategy devised for PDFs can be adapted to account for heavy quark mass effects in FFs extractions. Indeed, different schemes have already been applied to assess heavy quark hadronization probabilities into heavy flavored hadrons \cite{Cacciari:2005ry}. More recently, a GMVFN scheme, based on the FONLL scheme \cite{Forte:2010ta,Ball:2015tna} and aimed to improve the precision in parton-to-pion FFs, was successfully implemented in a QCD global analysis \cite{Epele:2016gup}. The mass dependent picture introduced by a GMVFN scheme was shown to modify significantly the hadronization probabilities of charm quarks into pions, due to the retention of the hard scattering mass effects in the partonic cross sections, rather than factorizing them into the FFs, in a consistent way. Additionally, the approach improves the quality of the fit to data, especially for data standing closer to the heavy quarks mass thresholds, and reduces the normalization shifts customarily included to accommodate data sets in the analysis \cite{deFlorian:2014xna}. 

In the following implement the GMVFN scheme in a global analysis designed to extract parton-to-kaon FFs. As it was done in the case of pion FFs, the dynamic effects related to the heavy quarks masses are computed to order $\alpha_\text{S}$ in perturbation theory for the single inclusive electron-positron annihilation (SIA) cross section. In the next section, we discuss very briefly kaon fragmentation and the role of heavy quarks in it. Then, we sketch how the GMVFN scheme is implemented within the complex Mellin moment technique in a global fit to current data in an efficient way, and show the corresponding results. We find a sizeable modification of the shape of the charm-to-kaon fragmentation function compared to the one obtained from the massless QCD approximation. The bottom fragmentation is found to be similar to the one obtained in the massless approximation, as it is constrained mostly by data well above the bottom mass production threshold, where mass effects are suppressed. At variance with the analysis performed using much more precise pion production data, we see only a mild improvement in the quality of the fit. 

\section{\label{sec:level2} Heavy quark into kaon fragmentation}
First efforts to determine light hadron FFs through next-to-leading order (NLO) QCD analyses \cite{ref:other-ffs,deFlorian:2007aj} typically focussed on very precise electron-positron annihilation data collected at LEP and SLAC, at energy scales close to the $Z$-boson mass, that is roughly twenty times larger than the bottom mass. To reproduce these data, a massless approach for the parton dynamics at first sight is a sensible approximation. Even extending SIA only studies to true global analyses, where extra information coming from hadron production in semi-inclusive deeply inelastic scattering (SIDIS) and in proton-proton (pp) collisions at much lower energy scales is used as a complement, the approximation seems still justified. In these additional processes, the contributions triggered by heavy flavor are strongly suppressed, hidden among many other contributions related to sea quarks in the initial state protons.

\begin{figure}
    \centering
    \includegraphics[width=\linewidth]{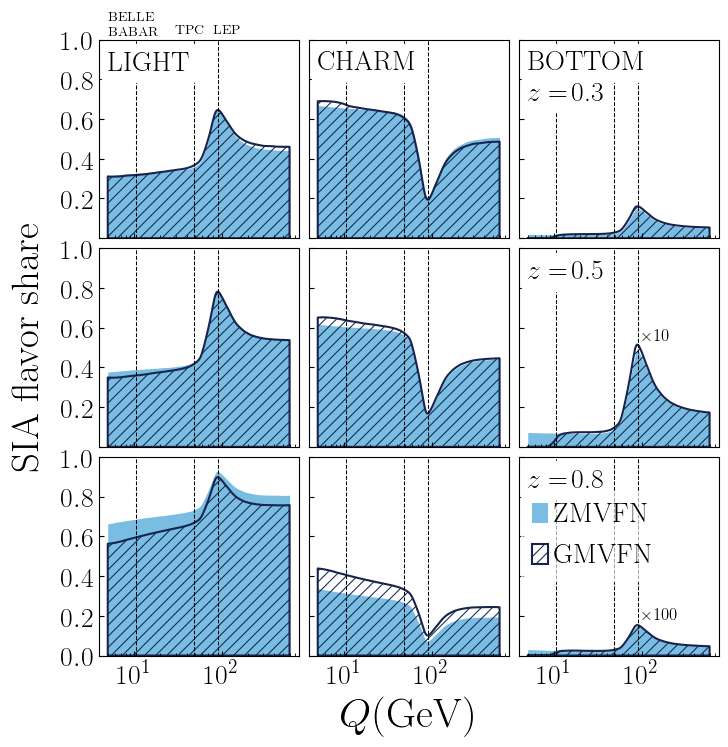}
    \caption{Comparison between the relative contributions of light, charm and bottom flavor 
    hadronization to the total single-kaon production cross section in electron-positron 
    annihilation process computed with the ZMVFN and GMVFN schemes.}
    \label{fig:cs-comp}
\end{figure}

Nevertheless, global analyses based on the massless perturbative QCD approximation, show that the nonperturbative hadronization probabilities for heavy quarks into light hadrons are, themselves, not negligible at all \cite{deFlorian:2007aj}. Indeed, they may be as large in size as valence quark hadronization probabilities, and they contribute to a significant fraction of the SIA cross section. In order to illustrate this point, in Fig. \ref{fig:cs-comp} we show the fraction of the charged kaon SIA cross section contributed by light, charm and bottom quark fragmentation, respectively, as a function of the energy scale and in three different hadron momentum fraction regions. The colored areas represent the results coming from a typical ZMVFN analysis like the DSS07 \cite{deFlorian:2007aj} or DSS17 \cite{deFlorian:2017lwf} sets, but with slightly updated inputs that will be specified in the next section. Below the $Z$-boson mass scale, and for kaons carrying a not very large fraction ($z < 0.6$) of the total available center of mass energy, charm quark fragmentation dominates the SIA cross section. This dominant charm contribution comes in part from the size of the charm FF itself, comparable to the total {\it strange} FF, but that enters the SIA cross section multiplied by a four times larger electroweak charge factor, and that is also three times bigger than the total {\it up} FF. Approaching the $Z$-boson mass scale, the electroweak charge suppress the relative charm contribution.  For increasing kaon energy fractions, one recovers a more intuitive picture, where {\it strange} and {\it up} quark fragmentation dominates the SIA cross section. On the other hand, the bottom FF is found to be much smaller than the one for charm, and combined with the also smaller charge factor, produces a very minor contribution. Anticipating the results from the GMVFN scheme, the dashed areas show the same but estimated with FFs obtained keeping mass effects, and will be discussed in detail later. 

In addition to the typical size of the charm quark contribution to the SIA cross section, if a set of SIA data sits close to the mass thresholds, then the mass effects are no longer suppressed and consequently are expected to become relevant in the flavor separation. Mass effects would have a direct impact in the charm quark FF through the SIA cross section coefficients and also an indirect effect in the gluon FF through its evolution to high energies, since it is coupled to the charm. In fact, in the last decade there has been a substantial improvement in the precision of hadron production measurements at relatively low energy scales such as Belle \cite{ref:belledata} and BaBar \cite{ref:babardata} experiments, rising the question of how necessary is it to improve the description of hadronization processes including the dynamics of heavy quarks to match their precision. 

\section{\label{sec:level2} GMVFN scheme global analysis}
The most simple way to account for the dynamical effects associated to the heavy quark masses consists in the implementation of factorization schemes with a fixed number of flavors (FFNS). In these schemes, the partonic cross sections are computed retaining the mass dependent terms and the number of active flavors is defined by the energy scale of the process to be described. These schemes are appropriate to reproduce hard interaction processes at energy scales close to the heavy flavor masses, however, they are not adequate to handle multiple energy scale problems, like a QCD global analysis. The limitation comes precisely from treating heavy quarks as massive particles always, what leads to some potentially dangerous logarithmic contributions in the partonic cross sections when the energy scale becomes much larger than the mass scales. Such logarithmic contributions may spoil the accuracy of perturbative calculations. 

In the opposite scenario, in the renormalization group improved massless quark approach, a process independent resummation of the logarithmic contributions deals with the problem at high energies, but gives an obviously inadequate description close to the mass thresholds. The general mass variable flavor number scheme (GMVFN) is designed to interpolate continuously and smoothly between the low energy regime, where the heavy quarks are treated as massive particles, and the high energy one, where the dynamic effects of all parton masses are negligible. In this way, the dynamics of heavy quarks are consistently described across the entire range of energy scales relevant for a QCD global analysis. As for any factorization scheme, the definition of GMVFN scheme is not unique. There is a certain degree of arbitrariness that reflexes in how fast the massive picture convergence to the massless limit \cite{Epele:2016gup}. This particular feature of the approach, rather than a weakness, can be exploited to optimize the description of data.

The implementation of a GMVFN scheme in a QCD global analysis for FFs has already been discussed in detail in the case of pion FFs in \cite{Epele:2016gup}. The main point is that appropriately subtracted massive cross sections are convoluted with the parametrizations of the corresponding fragmentation functions, evolved through the standard DGLAP evolution equations \cite{ref:dglap}. The unknown parameters, that define the hadronization probabilities, are determined as usual by comparing experimental measurements and their theoretical predictions, through a suitable  $\chi^2$ function: 
\begin{equation}
\label{eq:chi2}
\chi^2=\sum_{i=1}^N  \left[
\left( \frac{1-{\cal{N}}_i}{\delta{\cal{N}}_i}  \right)^2 +
\sum_{j=1}^{N_i} \frac{({\cal{N}}_i T_j-E_j)^2}{\delta E_j^2} \right],
\end{equation}
here $i=1,\ldots,N$ labels the data sets, each contributing with $N_i$ data points. $E_j$ is the measured value for a given observable, $\delta E_j$ the error associated with this measurement, and $T_j$ is the corresponding theoretical estimate for a given set of parameters. Since the full error correlation matrices are not available for some of the data sets used in the fit, statistical and systematical errors are simply added in quadrature in $\delta E_j$ as in previous fits \cite{deFlorian:2007aj,deFlorian:2014xna,deFlorian:2017lwf}. Normalization shifts ${\cal{N}}_i$, introduced for each data set to account for the quoted normalization uncertainties are computed analytically from the condition $\partial \chi^2/\partial {\cal{N}}_i=0$ in each iteration, with a corresponding penalty.

The Mellin moment approach is used to perform parameter determination in an efficient way \cite{deFlorian:2007aj}. This technique allows to replace every convolution integral by simple products. Because of the the complexity of the mass dependent expressions inherent to the GMVFN partonic cross sections, it is helpful to numerically compute their Mellin transforms and tabulate them before the global fit is performed. To recover the cross sections in the $z$-space, Mellin inversion integrals require the use of appropriate contours in the line integrals in complex moment space. As in ref. \cite{Epele:2016gup}, the specific prescription for the GMVFN scheme is chosen in order to optimize the description of the complete set of experimental data included in the global analysis. We found that the same prescription as in the case of pions, suggesting that the preference for a more faster or slower convergence to the massless limit could be universal, is related to the heavy quarks dynamics rather than to the final state hadron species.

The SIA-to-parton cross sections were computed at NLO in perturbation theory retaining charm quark and bottom quark masses in the framework of a GMVFN scheme. The value for the running strong coupling $\alpha_s$ is the one obtained in the NNPDF3.0 set of PDFs \cite{Ball:2014uwa}, which implements also a GMVFN factorization scheme. Hadroproduction cross sections in proton-proton collisions and SIDIS are computed with this PDF set, but in the massless parton approximation, since for these particular processes heavy quark contributions are strongly suppressed relative to the lighter flavors. The use of these PDFs has been shown to give a much better description of SIDIS data than other PDF sets  \cite{Borsa:2017vwy}.

\begin{table}[h!]
\caption{\label{tab:expkaontab} Individual $\chi^2$ contributions and normalization shifts $N_i$ for the data sets included in two global analyses where the ZMVFN and GMVFN schemes have been implemented.}  
\begin{ruledtabular}
\begin{tabular}{lcccccc}
%
%
experiment                           & data                 & \# data & \multicolumn{2}{c}{ZMVFN} &\multicolumn{2}{c}{GMVFN} \\
                                     & type                 & in fit  & $\mathcal{N}_i$ & $\chi^2$& $\mathcal{N}_i$ & $\chi^2$ \\
\hline \\[-7.5pt]
{\sc Aleph} \cite{ref:alephdata}     & incl.\               & 13      & 1.011       &   8.6   & 1.023       &   7.6    \\
{\sc BaBar} \cite{ref:babardata}     & incl.\               & 30      & 1.065       &  24.4   & 1.005       &  10.4    \\ 
{\sc Belle} \cite{ref:belledata}     & incl.\               & 78      & 0.983       &  16.5   & 1.009       &  14.7    \\  
{\sc Delphi} \cite{ref:delphidata}   & incl.\               & 12      & 1.000       &   7.8   & 1.000       &   5.3    \\
                                     & $uds$ tag            & 12      & 1.000       &   7.9   & 1.000       &   8.1    \\
                                     & $b$ tag              & 12      & 1.000       &   4.0   & 1.000       &   2.8    \\
{\sc Sld} \cite{ref:slddata}         & incl.\               & 18      & 1.002       &   8.0   & 1.005       &   7.4    \\
                                     & $uds$ tag            & 10      & 1.002       &  13.3   & 1.005       &  12.3    \\
                                     & $c$ tag              & 10      & 1.002       &  19.1   & 1.005       &  17.7    \\
                                     & $b$ tag              & 10      & 1.002       &  11.6   & 1.005       &  11.8    \\
{\sc Tpc} \cite{ref:tpcdata}         & incl.\ 34 GeV        &  4      & 1.000       &   1.8   & 1.000       &   1.9    \\
{\sc Tpc} \cite{ref:tpcdata}         & incl.\ 29 GeV        & 12      & 1.000       &  12.2   & 1.000       &   10.0    \\ 
\hline \\[-7.5pt]
{\sc Compass} \cite{ref:compassmult} & $K^{+}$(d)            & 309    & 1.012       & 229.2   & 1.013       &  229.4   \\
                                     & $K^{-}$(d)            & 309    & 1.012       & 211.8   & 1.013       &  209.5   \\
{\sc Hermes} \cite{ref:hermesmult}   & $K^{+}$(p) $Q^2$      & 36     & 0.830       &  62.6   & 0.832       &   61.2   \\
                                     & $K^{-}$(p) $Q^2$      & 36     & 0.830       &  34.2   & 0.832       &   34.0   \\
                                     & $K^{+}$(p) $x$        & 36     & 1.124       &  69.1   & 1.127       &   69.0   \\
                                     & $K^{-}$(p) $x$        & 36     & 1.124       &  35.1   & 1.127       &   35.7   \\
                                     & $K^{+}$(d) $Q^2$      & 36     & 0.836       &  41.8   & 0.838       &   41.0   \\
                                     & $K^{-}$(d) $Q^2$      & 36     & 0.836       &  36.2   & 0.838       &   35.7   \\
                                     & $K^{+}$(d) $x$        & 36     & 1.091       &  38.3   & 1.094       &   38.5   \\
                                     & $K^{-}$(d) $x$        & 36     & 1.091       &  32.0   & 1.094       &   32.2   \\ 
\hline \\[-7.5pt]
{\sc Star} \cite{ref:stardata13}     & $K^{+}$,$K^{+}/K^{-}$ & 16     & 1.085       &   7.6   & 1.085       &    7.5   \\
{\sc Alice} \cite{ref:alicedata}     & $K/\pi$               & 15     & 0.991       &  11.7   & 0.992       &   11.0   \\[2.5pt] 
\hline\hline \\[-5pt] 
{\bf TOTAL:}                         &                       & 1158   & \multicolumn{2}{c}{944.8}& \multicolumn{2}{c}{913.9} \\
\end{tabular}
\end{ruledtabular}
\end{table}

At variance with the pion SIA data, the kaon production cross section measured by {\scshape BaBar} collaboration at $10.54\text{ GeV}$, is about a 10\% larger than the one measured by the {\scshape Belle} experiment at $10.52\text{ GeV}$ in most of the kaon energy fraction range ($ 0.3 < z < 0.8$). This difference is significantly larger than the normalization uncertainties estimated by both collaborations. Different strategies to estimate and subtract kaons from secondary decays in both experiments, for example, could contribute to such differences. In any case, a full analysis of the origin of this feature is beyond the scope of the present analysis. In ref.\cite{deFlorian:2017lwf}, the apparent difference between both data sets was treated as a typical normalization error and absorbed into the normalization shifts ${\cal{N}}_i$ introduced in Eq.~(\ref{eq:chi2}). In consequence, the fit to data negotiate an intermediate solution that reproduce neither of the data sets. Alternatively, one could assume the difference between both experiments as coming from two differently defined observables, and let the minimization decide which definition suits best the rest of data in the fit, shifting the second so that both data sets agree. Doing this, we find a much better overall agreement between data and theory shifting down by around a 10\% {\scshape BaBar} data set. Another possibility would be to consider that the data sets are not compatible and eliminate one of them from the fit, retaining the one that leads to best overall agreement. However, we find that this last approach leads basically to the same result as in the previous alternative.

In Table \ref{tab:expkaontab} we present the results of two analyses implementing ZMVFN and the GMVFN schemes, respectively, indicating the partial contributions to $\chi^2$ and normalizations for each of the data sets included in the fit. As it can be noticed, most of the SIA data sets are slightly better described in the GMVFN framework and with typically smaller normalization shifts. The effect is particularly noticeable for {\scshape BaBar} and {\scshape Belle} with a reduction from 6.5 to 0.6 \%, and 1.7 to 0.9 \% respectively. Notice that the additional normalization applied to {\scshape BaBar} data discussed in the previous paragraph and motivated by a direct comparison between the two data sets at roughly the same center of mass energies, is implemented in both the ZMVFN and the GMVFN analyses. The improvement leading to smaller values for ${\cal{N}}_i$ is a direct consequence of the GMVFN scheme, since it induces a slightly different $z$-dependence in the SIA cross section that fits better the data, and further reduces the tension with other data sets at different energies.

\begin{figure}[b!]
    \centering
    \vspace{.25cm}
    \includegraphics[width=\linewidth]{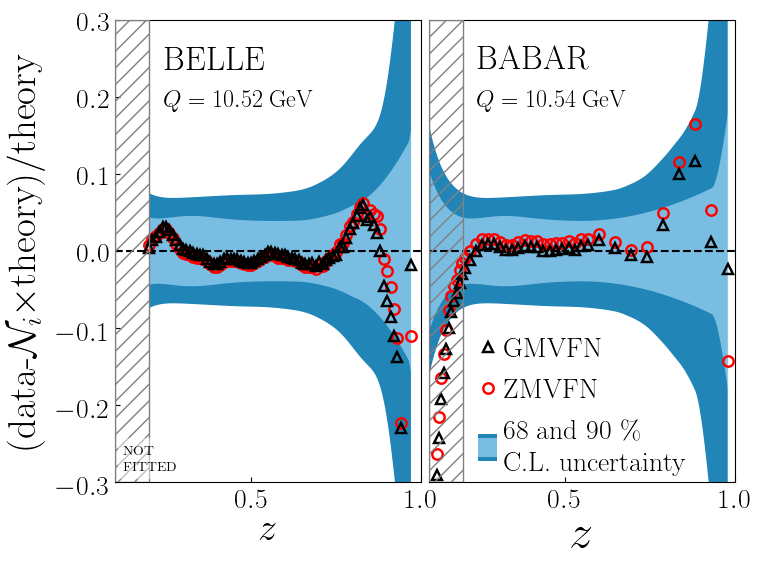}
    \caption{Comparison between data from {\sc Belle} and {\sc BaBar} and estimates from the ZMVFN and GMVFN schemes}
    \label{fig:bellbabar}
\end{figure}

\begin{figure}[t!]
    \centering
    \vspace{.25cm}
    \includegraphics[width=\linewidth]{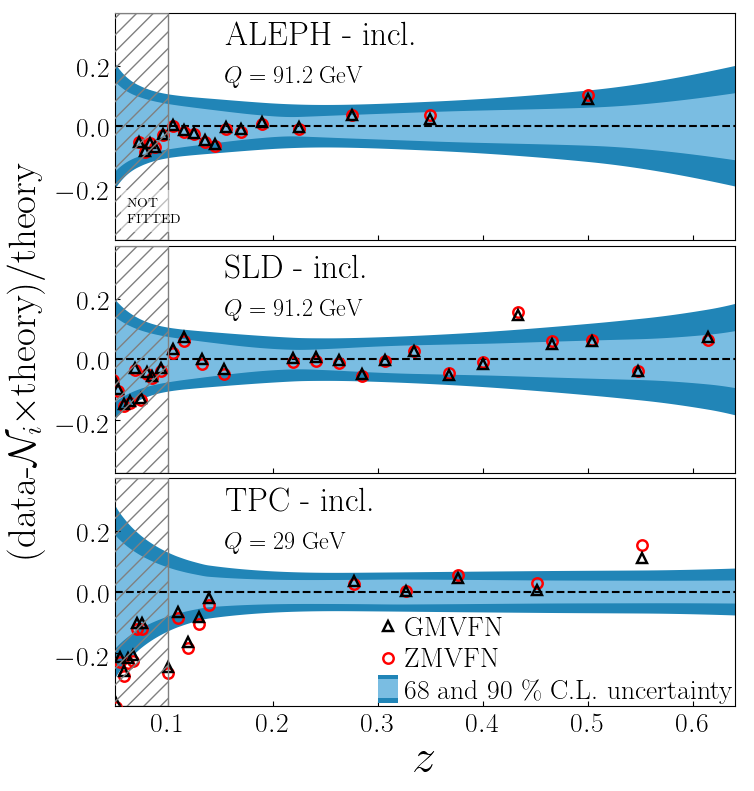}
    \caption{Comparison between data from ALEPH, SLD and TPC at higher energy scale and the corresponding ZMVFN and GMVFN schemes estimates. }
    \label{fig:zboson}
\end{figure}

In a previous QCD global analyses performed within the ZMVFN scheme \cite{deFlorian:2017lwf}, the exclusion of the bottom channel in the estimate of the  SIA kaon production cross section at low center of mass energy scale was necessary to reproduce {\scshape Belle} and {\scshape BaBar} measurements. In the GMVFN scheme, this contribution is highly suppressed near the $2m_b$ threshold, and in consequence the whole data included in the global fit is described in a much more natural way. 

Figs. \ref{fig:bellbabar} and \ref{fig:zboson} show the degree of agreement between SIA data sets at different energy scales and theory in both the GMVFN and the ZMVFN schemes. It is worth noticing that the improvement is not limited to the lower energy scale experiments but has an overall effect. Indeed, higher energy scale data sets, like {\scshape Aleph}, {\scshape Delphi} and {\scshape SLD} measurements, are also better described by the GMVFN scheme. As in ref.\cite{deFlorian:2017lwf} we include uncertainty estimates for the 68 and 90 \% confidence level limits, estimated with the improved hessian approach. 

\begin{figure}[t!]
    \centering
    \vspace{5pt}
    \includegraphics[width=\linewidth]{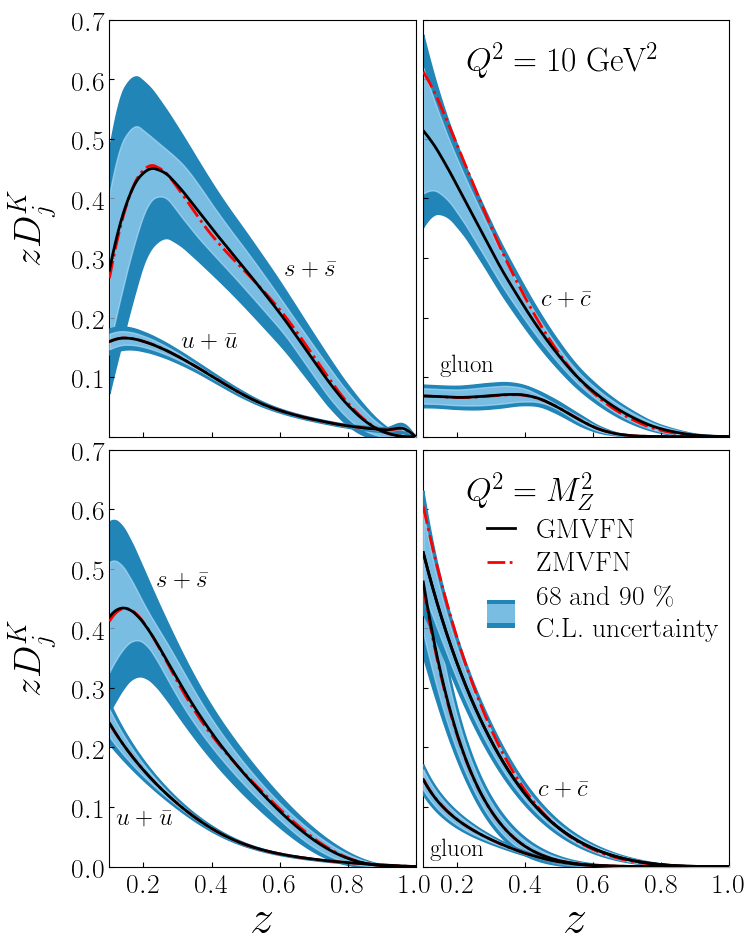}
    \caption{FFs at $Q^2 = 10$ GeV$^2$ and $Q^2 = M_Z^2$ coming from the ZMVFN and GMVFN schemes, respectively.}
    \label{fig:FFvsX}
\end{figure}

The resulting FFs are presented in Fig. \ref{fig:FFvsX} as a function of the momentum fraction $z$ for two different energy scales. As a consequence of the introduction of heavy quark mass effects, charm fragmentation probabilities are noticeably modified. The differences are larger than the numerical uncertainty bands computed with a 60\% C.L., for most of the $z$ values range, and are preserved at high energy scales by the evolution equations. It can be noticed that no significant differences are found for the light quark FFs. For these particular flavors, the hadronization probabilities are constrained mainly by light flavor tagged SIA and SIDIS data. Most of the high energy flavor tagged SIA data were acquired at the $Z$-boson mass energy scale, for which charm and bottom quark mass dependent corrections become negligible. On the other hand, {\sc Belle} and {\sc BaBar} constrain very little the bottom hadronization probabilities because of the suppression of its contributions to the SIA single kaon production in both the ZMVFN and the GMVFN schemes. As it shown in Fig. \ref{fig:FFvsX}, the bottom FF obtained from the inclusion of massive effects differs very little with the massless picture result.

Going back to the flavor share in the SIA cross section shown in Fig. \ref{fig:cs-comp}, but now computed with FFs extracted in GNVFN scheme, we can see to what extent the factorization of mass effects into the effective the ZMVFN scheme fragmentation functions leads to an inaccurate picture. As it can be noticed, there is a sizable difference between the estimates of the charm flavor role predicted by the ZMVFN and GMVFN schemes. The light flavor contribution estimated by the massless parton approximation are larger than the one obtained with a more consistent description of the heavy quarks dynamics. The difference between both schemes is more conspicuous as larger is the kaon energy fraction. The opposite is true for the charm contribution what implies that the charm contribution is typically underestimated in the massless framework. Finally, there is almost no difference between the massless and the massive subtracted predictions for the bottom quark contribution, respectively. Since the most important constraints to the bottom hadronization probabilities are provided by the highest energy data sets, bottom FFs extracted with both schemes are practically identical, and the difference between the partonic cross sections computed through the implementation of the GMVFN and the ZMVFN schemes is suppressed by the convolution with the both small respective FFs.

\section{Conclusions}
An extension of the GMVFN scheme to extract NLO FFs of quarks and gluons into kaons has been presented. Even though heavy quark mass effects put in evidence by this scheme are comparatively moderate, they already make a difference with present data and will certainly be required to match the precision of the future generation of hadroproduction experiments. The GMVFN framework induces a different energy scale dependence for the heavy quark contribution to the SIA cross section, together with a considerable suppression of these flavors near their mass thresholds. These features lead to inaccurate estimates of the relative importance between light and heavy flavor contributions to the leptoproduction of kaons. Specifically, light flavour contribution computed with the ZMVFN is typically larger than the GMVFN results. The differences is more noticeable when the final state kaons carry a larger fraction of the total available energy. This increase is balanced by the charm contribution, which shows the opposite behavior.


\end{document}